%% file: krakow.tex
\documentclass{appolb}
\usepackage{epsfig,amsmath,subfigure,latexsym,amssymb}
\def\lsi{\raise0.3ex\hbox{$<$\kern-0.75em\raise-1.1ex\hbox{$\sim$}}}
\def\gsi{\raise0.3ex\hbox{$>$\kern-0.75em\raise-1.1ex\hbox{$\sim$}}}

\newcommand{\gsim}{\mathop{\gsi}}
\newcommand{\be}{\begin{equation}}
\newcommand{\ee}{\end{equation}}
\newcommand{\bea}{\begin{eqnarray}}
\newcommand{\eea}{\end{eqnarray}}
\newcommand{\nn}{\nonumber}
\newcommand{\uno}{1 \!\! 1}
\newcommand{\Z}{\mathbb{Z}}
\newcommand{\einschub}{\small }

\newcommand{\la}{\langle}
\newcommand{\ra}{\rangle}

\begin{document}
\eqsec  
\title{The Non-Commutative $\lambda \phi^{4}$ Model %
\thanks{Talk presented by W.B. at Workshop on Random Geometry, Krakow, 2003}%
}
\author{W. Bietenholz, F. Hofheinz
\address{Institut f\"{u}r Physik, Humboldt Universit\"{a}t zu Berlin \\
Newtonstr. 15, D-12489 Berlin, Germany}
\and
J. Nishimura
\address{High Energy Accelerator Research Organization (KEK)\\
1-1 Oho, Tsukuba 305-0801, Japan}
}
\maketitle
\begin{abstract}
In the recent years, field theory on non-commutative (NC) spaces
has attracted a lot of attention. Most literature on this subject deals
with perturbation theory, although the latter runs into grave
problems beyond one loop. Here we present results from a fully
non-perturbative approach. In particular, we performed numerical
simulations of the $\lambda \phi^{4}$ model with two NC spatial
coordinates, and a commutative Euclidean time. This theory is lattice
discretized and then mapped onto a matrix model. The simulation results
reveal a phase diagram with various types of ordered phases. We discuss
the suitable order parameters, as well as the spatial and temporal 
correlators. The dispersion relation clearly shows a trend towards
the expected IR singularity. Its parameterization provides the tool to
extract the continuum limit.

\end{abstract}

Preprint {\bf HU-EP-03/69}


\section{Field Theory on a Non-Commutative Space}

The simplest way to introduce non-commutative (NC) coordinates
is to impose the relation
\be
[\hat x_{\mu}, \hat x_{\nu} ] = i \Theta_{\mu \nu} \ ,
\ee
$\Theta$ being a constant, anti-symmetric tensor, while $\hat x_{\mu}$
are Hermitian coordinate operators (which cannot be diagonalized 
simultaneously). This relation leads to an uncertainty in c-space,
$\Delta x_{\mu} \Delta x_{\nu} > 0$ $(\mu \neq \nu )$.
The pre-history of this idea involves private communication among
celebrities like Heisenberg, Peierls, Pauli and Oppenheimer.
The latter asked his student Snyder to work it out, which yielded the
first publication about physics on a NC space \cite{Sny},
followed immediately by Ref.\ \cite{CNY}.

The mathematical framework for the formulation of {\em field theory on NC 
spaces} was worked out in the eighties \cite{Connes}. However, a real
boom of interest was triggered only in the late nineties by the observation
that string and M theory at low energy in a magnetic background field can be
identified with NC field theory \cite{string}. 
\footnote{Strictly speaking, also that observation occurred
in the literature much earlier \cite{ACNY}.}
This boom persists up to now,
as is manifest from new preprints on this subjects appearing day after day.

The identification of Refs.\ \cite{ACNY,string}
transforms the background field into a $\Theta$ term. 
The same idea 
has been around in solid state physics long before, where it led
to a new description of the quantum Hall effect \cite{Girvin}.
The origin of this idea goes back to Peierls: an electron in a strong
magnetic background field has an obvious description by NC coordinates,
where the magnetic field is replaced by non-commutativity,
with an extent inverse to the magnetic field strength.\\

{\einschub 
To illustrate this fundamental property in the simplest possible way,
we consider an electron moving in a plane, with position
$\vec x = (x_{1},x_{2},0)$, exposed to a magnetic field 
$\vec B = (0,0,B)$. If this field is strong, the Lagrangian
\be
L[\vec x, \dot {\vec x}] = 
\frac{m}{2} \dot {\vec x}^{\, 2} + e B \epsilon_{ij} x_{i} \dot x_{j} 
\ , \qquad (i,j = 1,2)
\ee
can be reduced to the second term, so that the canonical momentum reads
\be  \label{canon}
\Pi_{j} = \frac{\partial L}{\partial \dot x_{j}} 
= e B \epsilon_{ij} x_{i} \ .
\ee
Applying now the canonical quantization rule
\be
[ \hat x_{i}, \hat \Pi_{j} ] = i \hbar \delta_{ij}
\ee
we arrive at
\be
[ \hat x_{i}, \hat \Pi_{i} ] = 
e B \epsilon_{ij} [\hat x_{i},\hat x_{j}] \ .
\ee
Indeed, together with eq.\ (\ref{canon}) this corresponds to the NC relation
\be
 [\hat x_{i},\hat x_{j}] = i \Theta_{ij} \ , \qquad
\Theta_{ij} = \frac{\hbar}{e} \frac{1}{B} \epsilon_{ij}
:= \theta \epsilon_{ij} \ .
\ee
This illustration of Peierls' map, along with a more precise
derivation based on the Hamiltonian formalism, is explained for
instance in Ref.\ \cite{Barbon}.}\\

Hence one motivation to study NC field theory is simply its
application as a formalism to describe certain effects
in the commutative world. Such effects are typically related to
a background field, which is then transformed away by going
non-commutative.

However,
in addition to that concept, the present fashion also includes
studying the possibility of a really NC space-time. A deep, qualitative
difference from the commutative space-time is the occurrence of a 
{\em non-locality} of range $\sqrt{ \Vert \Theta \Vert }$.
Obviously this feature raises conceptual problems, but from the optimistic
point of view it is a source of hope for a link to quantum gravity.

In fact, there is a claim that attempts to
merge quantum theory and gravitation imply quite generally
a NC space. To illustrate this line of thought, we quote a simple
{\em Gedankenexperiment}. \\

{\einschub 
Some event is measured with accuracy $\Delta x$, $\Delta y$, $\Delta z$,
$\Delta t$. This requires an energy concentration, which implies a 
gravitational field. In the extreme case, the latter imposes an event
horizon beyond the uncertainty, so that the event is effectively invisible.
One may now evaluate the condition for avoiding this, i.e.\ for dealing
with actually detectable events.  On the Planck scale, an estimate in
Ref.\ \cite{DFR} suggests the constraints
\bea
\Delta x \Delta y + \Delta x \Delta z + \Delta y \Delta z \geq 1 
\ , \nn \\
(\Delta x + \Delta y + \Delta z ) \Delta t \geq 1 \ , 
\eea
so that the NC space seems indeed natural as soon as gravity is
involved.

However, this argument should come along with at least one remark 
of caution: much of the literature excludes time from the
non-commutativity, including the remainder of this article.
Otherwise the problems related to causality \cite{causal} are especially
severe. From the above argument, however, this step would not
be justified.} \\

The last point is also related to the question if and how the Wick rotation
from an Euclidean to a Minkowski signature can be performed in the NC
world. This is not ultimately settled yet, but in our case of
a commutative time the issue is less problematic.
Anyhow, here we work in Euclidean space, and we are happy there, without
worrying about the details of the transition to the NC Minkowski space.\\

Since the idea that our space could really be NC is fashionable,
of course there are already numerous speculations about possible
measurements of $\Theta$. \\

{\einschub One suggestion is based on the deformation
of the photon dispersion relation due to $\Theta$ \cite{Kamel}. 
Blazers (highly active galactic nuclei) emit bursts of photons over 
a broad energy spectrum.
Assuming this emission to be simultaneous, a relative delay could in
principle establish bounds on $\Theta$. However, in addition to
experimental difficulties, the knowledge
about the deformed dispersion is also limited to a one loop
calculation \cite{photon}.} \\

The last limitation is especially worrisome in the light of the
fact that most higher loop calculations are not feasible yet ---
no systematic machine is known for them. Perturbation theory is even more 
complicated than in the good old commutative space, which is in striking
contradiction to the original hope that $\Theta$ would simplify
the perturbative renormalization \cite{Sny}. It is true that part of the UV 
singularities are removed due to the non-commutativity, but others
remain, in particular those in the planar diagrams \cite{Filk}.
What makes the situation worse is that the non-planar divergences
do not just disappear, but they are rather turned into IR singularities
with respect to external momenta.
This effect is denoted as {\em UV/IR mixing} \cite{MRS}.
At this point we want to give again just a simple intuitive reason;
we will be somewhat more explicit below in the framework
of the $\lambda \phi^{4}$ model.\\

{\einschub
We return to simple natural units $\hbar =1$, without
involving the Planck scale any more. If we combine Heisenberg's 
uncertainty $\Delta x_{j} \sim 1/ \Delta p_{j}$ with the NC relation
\be
\Delta x_{j} \sim \Theta_{ij} / \Delta x_{i} \sim \Theta_{ij}
\Delta p_{i} \ , \qquad (i \neq j)
\ee
we see that for $\Delta p_{j} \to 0$ the other momentum components
explode, $\Delta p_{i} \to \infty$, and vice versa. This also suggests
that in addition to the Heisenberg term, $\Delta x_{j}$ picks up
a term linear in the $\Delta p_{i}$, which may be denoted as
a ``string modification'' of the uncertainty principle.}\\

In the work to be presented here we are going to consider a 3d Euclidean 
space with a commutative Euclidean time $t$ and two NC spatial coordinates,
which obey
\be
[ \hat x_{\mu} , \hat x_{\nu} ] = i \theta \epsilon_{\mu \nu} \ .
\ee
We are happy to avoid the horror of NC perturbation theory by taking
the {\em fully non-perturbative approach of numerical simulations}.
To this end, the space should first be {\em lattice discretized},
as in the commutative world. Here we also need a second step,
namely a mapping of the lattice field theory onto a {\em matrix model},
which will be described in Section 2.
At this point we only comment on the general structure of a
2d NC lattice of spacing $a$, following Ref.\ \cite{Szabo}.\\

{\einschub
The restriction of the spectrum of $\hat x_{\mu}$ to the lattice
sites corresponds to the operator identity
\be  \label{NClat}
\exp \Big( i \frac{2\pi}{a} \hat x_{\mu} \Big) = \hat \uno \ .
\ee
As in the commutative case, we want the momentum components $k_{\mu}$
to be periodic over the Brillouin zone,
\be
\exp \left( i \Big[ k_{\mu} + \frac{2\pi}{a}\Big] \hat x_{\mu} \right) 
= \exp ( i k_{\mu} \hat x_{\mu} ) \ .
\ee
Multiplying both sides by the factor $\exp (-i k_{\nu} \hat x_{\nu})$
now leads to consistency with eq.\ (\ref{NClat}), {\em iff}
\be
\frac{\theta}{2a} k_{\mu} \in \mathbb{Z} \ ,
\ee
which has amazing consequences: any NC lattice is automatically periodic,
say over a lattice volume $L \times L$. Then the discrete momenta
$k^{(n)} = \frac{2\pi}{aL} n$ occur, where $n_{\mu} \in \Z$.
The non-commutativity parameter can be identified as
\be  \label{theta-a-L}
\theta = \frac{1}{\pi} a^{2} L \ .
\ee 

We see that the continuum limit $a \to 0$ and the thermodynamic
limit $L \to \infty$ are manifestly entangled, which is
again an aspect of UV/IR mixing. Taking these two limits simultaneously
in such a way that $\theta$ remains constant is denoted as the {\em
double scaling limit}.
}

\section{The non-commutative $\lambda \phi^{4}$ model}

NC field theories can be formulated in a form which looks
similar to the commutative world, if all the fields are
multiplied by the {\em star product} (or {\em Moyal product}),
\be
f(x) \star g(x) := \exp \left( \frac{1}{2} i \Theta_{\mu \nu} 
\frac{\partial}{\partial x_{\mu}} \frac{\partial}{\partial y_{\nu}}
\right) f(x) \, g(y) \vert_{x=y} \ .
\ee
In the particular case of bilinear terms in an action, the star
product is equivalent to the ordinary product, hence in these terms 
the star product is not needed.

Based on these rules, we can write down for instance the action of
the NC $\lambda \phi^{4}$ model,
\be  \label{act4}
S [ \phi ] = \int d^{d}x \, \Big[ \frac{1}{2} \partial_{\mu} \phi
\, \partial_{\mu} \phi + \frac{m^{2}}{2} \phi^{2} + \frac{\lambda}{4}
\phi \star \phi \star \phi \star \phi \Big] \ .
\ee
Since only the self-interaction term involves $\Theta$, the coupling 
strength $\lambda$ also determines the extent of NC effects in this 
model. \\

{\einschub
To render the above star product rules plausible, we consider
the composition of plane wave operators,
\be
e^{i p_{\mu} \hat x_{\mu}} \cdot e^{i q_{\nu} \hat x_{\nu}} =
\exp \Big( - \frac{i}{2} p_{\mu} \Theta_{\mu \nu} q_{\nu} \Big)
\, e^{i (p+q)_{\mu} \hat x_{\mu}} \ .
\ee
This provides a prescription for the translation from the
commutative product $e^{ipx} e^{iqx}$ into a NC formulation in terms
of $x$, i.e.\ without using the operator $\hat x$,
\be
e^{ipx} e^{iqx} \to e^{-\frac{i}{2} p_{\mu} \Theta_{\mu \nu} q_{\nu}}
e^{i (p+q) x} := e^{ipx} \star e^{iqx} \ .
\ee
More generally, we can adopt this translation rule to the product of
fields, which can be decomposed into plane waves,
\be
\phi (x) \psi (x) \to \phi (x) \, e^{\frac{i}{2} 
\overset{\gets}{{\partial_{\mu}}} \Theta_{\mu \nu} 
\overset{\to}{{\partial_{\nu}}} } \, \psi (x) 
:= \phi (x) \star \psi (x) \ .
\ee

Turning now to the peculiarity of bilinear terms, it is easy to
see that in
\be
\int d^{d}x \, \phi (x) \star \psi (x) = 
\int d^{d}x \, \phi (x) \Big[ 1 - \frac{i}{2} 
\overset{\gets}{{\partial_{\mu}}} \Theta_{\mu \nu} 
\overset{\to}{{\partial_{\nu}}} + \dots \Big] \psi (x)
\overset{!}{=} \int d^{d}x \, \phi (x) \psi (x) \nn
\ee 
all the terms in the square bracket  which involve $\Theta$ cannot
contribute, based on partial integration and 
$\Theta_{\mu \nu} = - \Theta_{\nu \mu}$. } \\

The one loop level of this model is suitable for the illustration of
the general properties of NC perturbation theory that we mentioned
in Section 1.\\

{\einschub
To this end, we consider the one loop level of the one particle
irreducible 2-point function to the action (\ref{act4}),
i.e.\ the $n=1$ contribution to
\be  \label{2point}
\langle \phi (p) \phi (-p) \rangle = \sum_{n=0}^{\infty}
\lambda^{n} \Gamma^{(n)} (p) \ .
\ee
It contains a planar and a non-planar term,
\be  \label{2point1loop}
\Gamma_{planar}^{(1)} = \frac{1}{3}
\int \frac{d^{d}k}{(2\pi )^{d}} \frac{1}{k^{2} + m^{2}} \ , \quad
\Gamma_{non-planar}^{(1)} = \frac{1}{6}
\int \frac{d^{d}k}{(2\pi )^{d}} 
\frac{\exp (i k_{\mu} \Theta_{\mu \nu} p_{\nu})}{k^{2} + m^{2}} \ ,
\ee
which are illustrated in Fig.\ \ref{UVIRfig}.
\input{diagram}
We see that the planar part is independent of $\Theta$ \cite{Filk},
whereas the non-planar part is affected by UV/IR mixing \cite{MRS}.
To reveal what this means, we
introduce a momentum cut-off $\Lambda$ and obtain \cite{Szabo}
( for $m > 0$)
\be
\Gamma_{non-planar}^{(1)} (p) = \frac{m^{(d-2)/2}}{6 \, (2\pi )^{d/2}}
\Big( \frac{4}{\Lambda^{2}} - p_{\mu} \Theta^{2}_{\mu \nu} p_{\nu} \Big)
^{(2-d)/4} K_{\frac{d-2}{2}} \Big(
m \sqrt{ \frac{4}{\Lambda^{2}} - p_{\mu} \Theta^{2}_{\mu \nu} p_{\nu} }
\Big) \ ,  \nn
\ee
where $\Theta^{2}_{\mu \nu} = \Theta_{\mu \rho} \Theta_{\rho \nu}$
and $K$ is the modified Bessel function. In particular in $d=4$ the 
divergent part is given by
\bea
\Gamma_{non-planar}^{(1)} (p) &=& \frac{1}{96 \pi^{2}}
\left( \Lambda_{\rm eff}^{2} - m^{2} \ln
\Big( \frac{\Lambda_{\rm eff}^{2}}{m^{2}} \Big)
\right) + {\rm finite~terms} \ , \nn \\
\Lambda_{\rm eff}^{2} &=& \frac{1}{\frac{1}{\Lambda^{2}} - p_{\mu}
\Theta^{2}_{\mu \nu} p_{\nu}} \ .  
\eea
In general the effective cut-off $\Lambda_{\rm eff}$ remains finite
in the UV limit $\Lambda \to \infty$, {\em but} it diverges if we take
simultaneously the IR limit $p \to 0$.} \\

A comprehensive one loop study of the NC $\lambda \phi^{4}$ model in
various dimensions has been performed in Ref.\ \cite{GubSon}.
That work dealt with a self-consistent Hartree-Fock type
approximation, which would be exact for the $O(N)$ model at
large $N$. The authors assumed its validity also for the $N=1$
theory and conjectured in particular a prediction for the
phase diagram. As a general feature, the system undergoes some ordering
at $m^{2} \ll 0$, which corresponds in some sense to a very low
temperature. If $m^{2}$ is lowered to that point, Gubser and Sondhi
predict in $d=3$ and $d=4$ the following behavior:
\begin{itemize}

\item At small $\theta$, there is an Ising transition to a uniform order,
as in the commutative case.

\item At larger $\theta$, the ordered state has a structure of stripes
or even more complicated patterns.

\end{itemize}

Note that the formation of such a stripe order implies the
spontaneous breaking of translation invariance. Therefore Gubser
and Sondhi did not expect this phase in $d=2$. We will return to
this point at the end of Section 3.

The same question was also studied by means of a renormalization group
analysis \cite{ChenWu}.\\

Our goal was a non-perturbative, quantitative verification of that
qualitative conjecture, and our methods are
Monte Carlo simulations. The lattice formulation
(c.f.\ Section 1) is not hard to write down, but it is very hard
to simulate due to the star product. The way out which enabled
efficient simulations is a mapping onto a {\em dimensionally reduced
matrix model}, as suggested in Ref.\ \cite{AMNS}. This method is a 
refinement of the matrix model approach to NC gauge theory in the 
continuum \cite{AIIKKT}.

Assume the field $\phi (\vec x, t)$ to be defined on a $N^{2} \times T$
lattice of unit spacing.
According to Ref.\ \cite{AMNS} the following action is
equivalent to the lattice action:
\bea
S [ \bar \phi ] &=& N \, {\rm Tr} \sum_{t=1}^{T} \Big[
\frac{1}{2} \sum_{\mu =1}^{2} \Big( \Gamma_{\mu}
\bar \phi (t) \Gamma_{\mu}^{\dagger} - \bar \phi (t) \Big)^{2} \nn \\
&& + \frac{1}{2} \Big( \bar \phi (t+1) - \bar \phi (t) \Big)^{2}
+ \frac{m^{2}}{2} \bar \phi^{2}(t) + \frac{\lambda}{4}
\bar \phi^{4} (t) \Big] \ .
\eea
Here $\bar \phi (t)$ ($t=1 \dots T$) are Hermitian $N \times N$
matrices. In the (commutative) time direction the kinetic term takes
the ordinary discrete form. The quartic term --- which is in the 
original formulation plagued by repeated star products --- looks
relatively simple now, although it should be noted that these
matrices are not sparse (in contrast to typical lattice
formulations of commutative field theory), hence the fourth power
requires some computation. However, much of the complication due
to the NC geometry is now manifest in the non-standard kinetic term
in the spatial directions. The so-called {\em twist eaters} 
$\Gamma_{\mu}$ provide the shift which corresponds to one lattice unit,
if they obey the 't Hooft algebra
\be
\Gamma_{\mu} \Gamma_{\nu} = {\cal Z}_{\nu \mu} \Gamma_{\nu} \Gamma_{\mu} \ .
\ee
The tensor ${\cal Z}_{\mu \nu} = {\cal Z}_{\nu \mu}^{*}$ is called the
{\em twist} factor. In general it may have the form
${\cal Z}_{12}= e^{2 \pi i k/N}$, where $k \in \Z$.
\footnote{Its standard form, which occurs for instance in the ``twisted
Eguchi-Kawai Model'' \cite{TEK}, uses $k=1$.}
The representation of the twist eaters that we choose \cite{BHN,diss} 
requires the twist to have the specific form
\be  \label{twist}
{\cal Z}_{12} = \exp \Big( i \frac{N+1}{N} \pi \Big) \ ,
\ee
as we are going to sketch below. Comparison with the general form shows
that then $N$ has to be odd. The lattice model and the matrix
model are connected by {\em Morita equivalence}, which means that their
algebras are fully identical.\\

{\einschub
Let us go back to the NC lattice formulation discussed in
Section 1. Since the momenta are discrete, we should not use the
(unbounded) operators $\hat x_{\mu}$, $\mu =1,2$ to describe the lattice 
sites. Instead we introduce the unitary operators
$ \hat Z_{\mu} = \exp ( \frac{2\pi}{La} i \hat x_{\mu})$, which obey the
commutation relation
\be  \label{con1}
\hat Z_{\mu} \hat Z_{\nu} = \exp\Big( -\frac{4 \pi^{2}}{a^{2}L^{2}} 
i \Theta_{\mu \nu} \Big) \hat Z_{\nu} \hat Z_{\mu} =
\exp \Big( -\frac{4 \pi}{L} i \epsilon_{\mu \nu} \Big)
\hat Z_{\nu} \hat Z_{\mu} \ ,
\ee
where we used $\Theta_{\mu \nu} = \theta \epsilon_{\mu \nu}$ and
eq.\ (\ref{theta-a-L}). The lattice translation operator
$\hat D_{\mu} = \exp (a \hat \partial_{\mu})$ is supposed to
fulfill the relation
\be  \label{con2}
\hat D_{\mu} \hat Z_{\nu} \hat D_{\mu}^{\dagger} =
 e^{2 \pi a i \delta_{\mu \nu} /L} \hat Z_{\nu} \ .
\ee
The issue is now to find a matrix solution for the conditions
(\ref{con1}) and (\ref{con2}). This solution is unique only
up to symmetry transformations,
hence one may end up with different twist factors. We assumed the 
unitary twist eaters to take the form
\be
\Gamma_{1} = \left( \begin{array}{cccccc}
0 & 1 & 0 & . & . & . \\
0 & . & 1 & . & . & . \\
. & . & . & 1 & . & . \\
. & . & . & . & . & . \\
. & . & . & . & . & 1 \\
1 & . & . & . & . & 0 \end{array} \right) \quad , \quad
\Gamma_{2} = \left( \begin{array}{cccccc}
1 & 0 & . & . & . & ~. \\
0 & {\cal Z}_{21} & 0 & . & . & ~. \\
0 & 0 & {\cal Z}_{21}^{2} & . & . & ~. \\
. & . & . & {\cal Z}_{21}^{3} & . & ~. \\
. & . & . & . & . & ~. \\
0 & . & . & . & . & ~. \end{array} \right) \nn
\ee
where ${\cal Z}_{12} = {\cal Z}_{21}^{*}$ characterizes the twist.
This ansatz solves indeed the conditions (\ref{con1}) and (\ref{con2}),
iff we choose the twist of eq.\ (\ref{twist}) \cite{diss}.
However, a different ansatz for $\hat D_{\mu}$ resp.\ $\hat Z_{\mu}$
may lead to an alternative solution with a modified twist, see
e.g.\ Ref.\ \cite{aristocats}.}

\section{Numerical results}

\begin{figure}[htbp]
 \begin{center} 
   \vspace{-5mm}
   \includegraphics[width=0.7\linewidth]{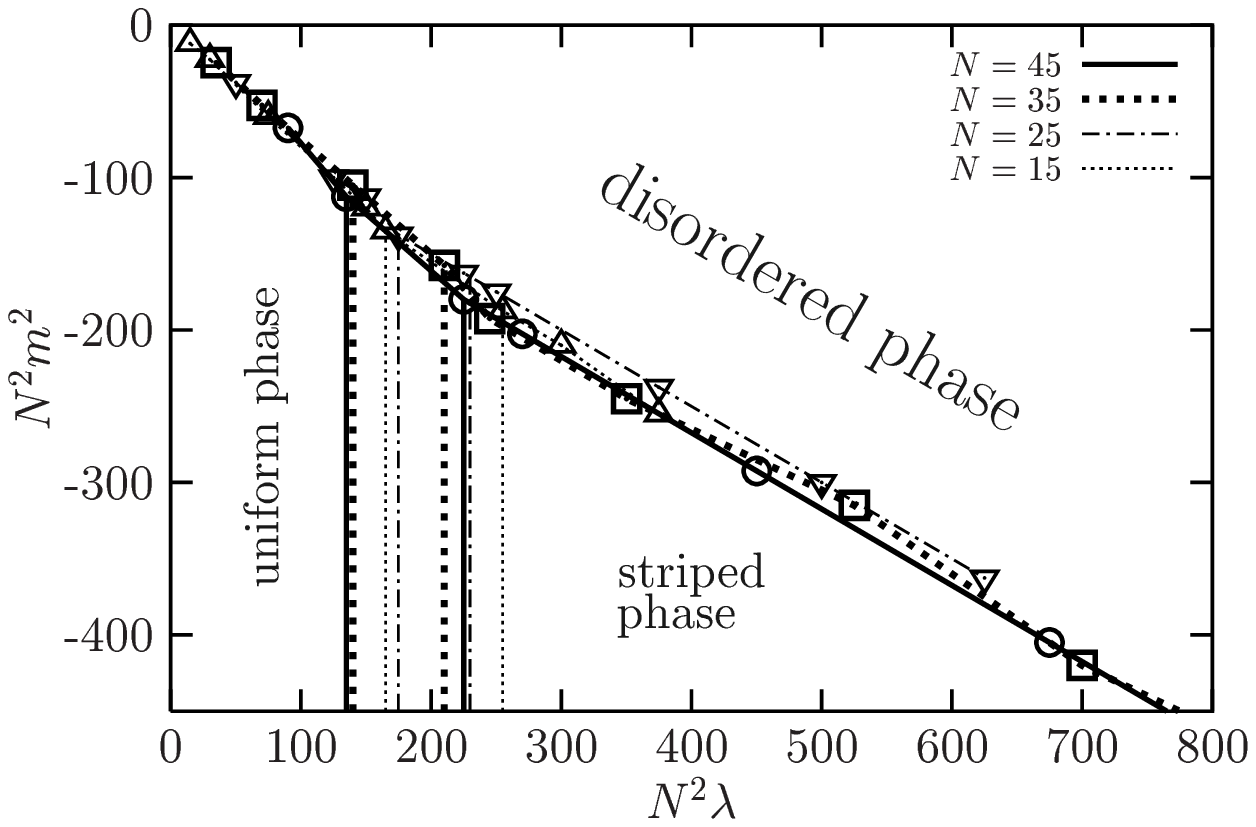} \\
 \end{center}
\caption{{\it The phase diagram of the NC $\lambda \phi^{4}$ model,
explored by means of numerical simulations.
Since it contains a new type of phase we conclude:
this diagram of phases simply amazes.}}
\label{phase-dia}
\end{figure}

\begin{figure}[htbp]
 \centering 
\subfigure[\footnotesize{$N=45,~\lambda = 0.044,~m^{2}=-0.011$}]
{\epsfig{figure=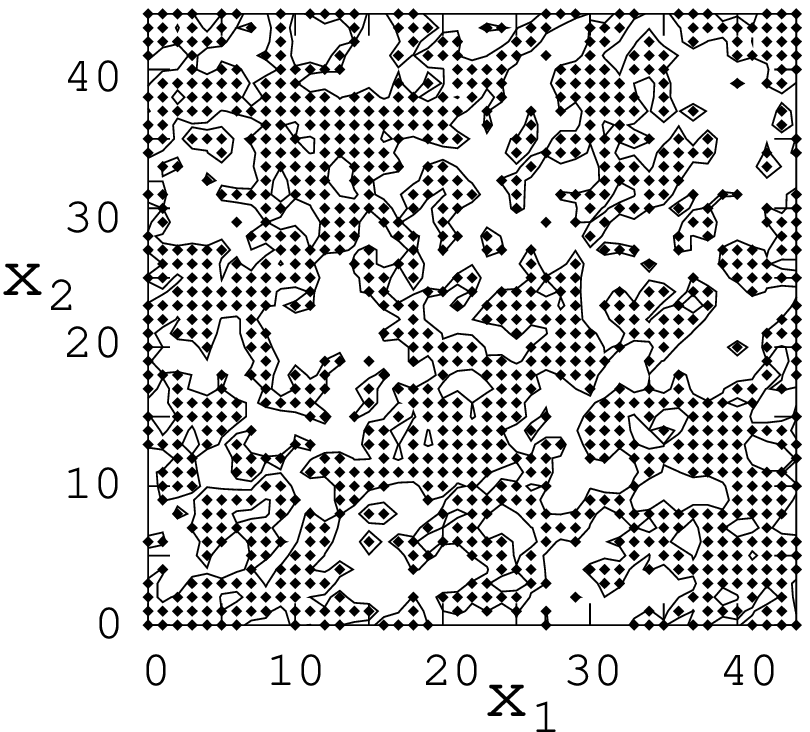,width=.3\linewidth}}%
\hspace*{1cm}
\subfigure[\footnotesize{$N = 45, ~ \lambda = 0.044,~m^{2}=-0.11$}]
{\epsfig{figure=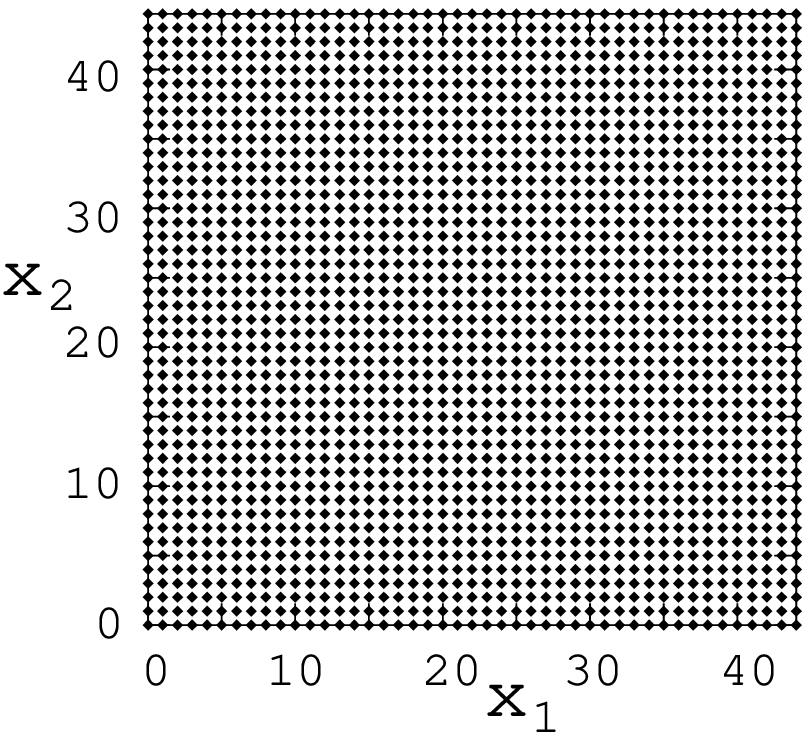,width=.3\linewidth}}
\\
\vspace*{-4mm}
  \subfigure[\footnotesize{$N=45,~\lambda = 0.22,~m^{2}=-0.2$}]
{\epsfig{figure=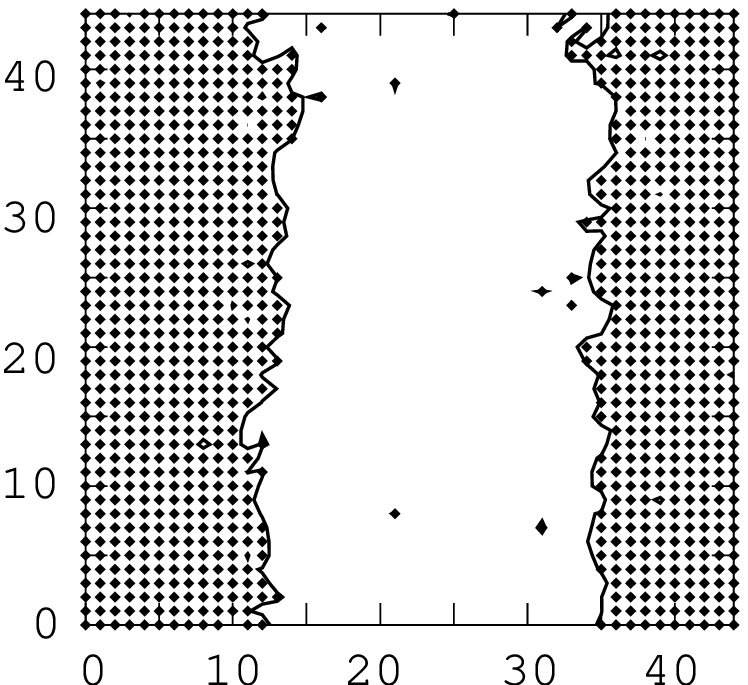,width=.28\linewidth}}%
\hspace*{6mm}
\subfigure[\footnotesize{$N=35,~\lambda = 10,~m^{2}=-4$}]
{\epsfig{figure=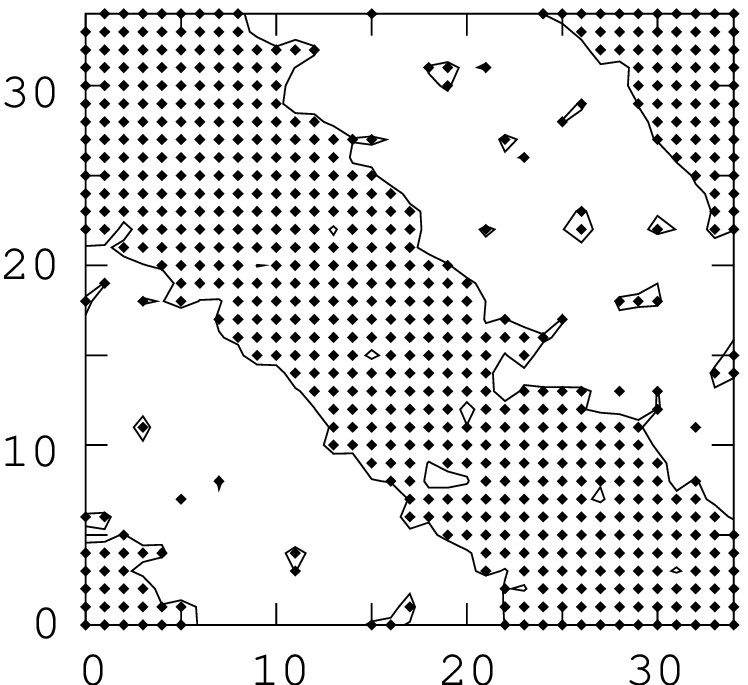,width=.28\linewidth}}
\hspace*{2mm}
\subfigure[\footnotesize{$N=55,~\lambda = 50,~m^{2}=-22$}]
{\epsfig{figure=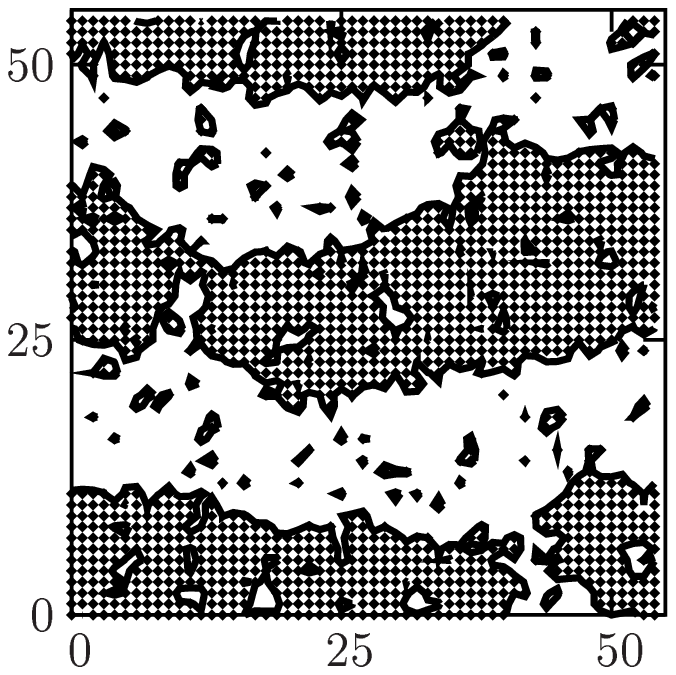,width=.27\linewidth}}
\caption{{\it Snapshots of typical, well thermalized configurations
in different sectors of the phase diagram: disordered and uniformly ordered 
(above), and patterns of two stripes parallel to an axis, two diagonal 
stripes and finally four stripes (below). The
dark and bright areas correspond to $\phi >0$, $\phi <0$.}}
\label{snap}
\vspace*{-5mm}
\end{figure}

An overview of our numerical results is given by the phase diagram
in Fig.\ \ref{phase-dia}. For a suitable re-scaling of the axis,
this diagram stabilizes with increasing $N = T$.
At strongly negative $m^{2}$ (corresponding to a very low temperature)
the system seeks some order.
At small self-interaction $\lambda$ the non-commutativity is not
much amplified and the systems behaves as in the commutative world,
i.e.\ it is ordered uniformly. At larger $\lambda$ a new {\em striped
phase} sets in, in the spirit of Ref.\ \cite{GubSon}. Such a phase
is not known in the commutative $\lambda \phi^{4}$ model, but
similar phenomena appear in solid state physics, see for instance
Ref.\ \cite{solid}.
In that phase, a non-zero mode condenses, so that the ground states
correspond to a stripe pattern, or to more complicated 
checker-field-type patterns. Typical examples for the various cases
are shown in Fig.\ \ref{snap}.

The transition from disorder to order can be localized well
and it appears to be second order. On the other hand, the transition
between the uniform and the striped phase is more difficult
to localize and it is expected to be of first order.

The order parameter that was used here is based on the spatial Fourier
transform of the scalar field, $\tilde \phi (\vec p , t)$. We average 
over the time direction and turn it such that an eventually condensated
mode can be detected optimally. This is done be introducing the function
\be
M(k) = \frac{1}{NT} \ ^{\rm max}_{k = \vert \vec p \vert N /2\pi} \,
\left| \sum_{t=1}^{T} \tilde \phi (\vec p , t) \right| \ ,
\ee
and its expectation value is our order parameter,
sensitive to the mode $k$. In particular, $\la M(0) \ra$ is the
standard order parameter for the $Z_2$ symmetry, $\la M(1) \ra$
is the staggered order parameter that detects patterns of two stripes
parallel to one of the axes
as in Fig.\ \ref{snap} below on the left, $\la M(\sqrt{2}) \ra$
detects a diagonal two-stripe pattern as in Fig.\ \ref{snap} below
in the center, $\la M(2) \ra$
detects a pattern of four stripes parallel to one of the axes
as in Fig.\ \ref{snap} below on the right, etc.
We also measured the connected two-point functions of $M(k)$, the peak
of which allowed us to localize the order--disorder
phase transition in the diagram
of Fig.\ \ref{phase-dia} to a high precision.

\begin{figure}[htbp]
 \centering
  \subfigure[\footnotesize{$\lambda = 0.06,~m^{2}=-0.015$}]
{\epsfig{figure=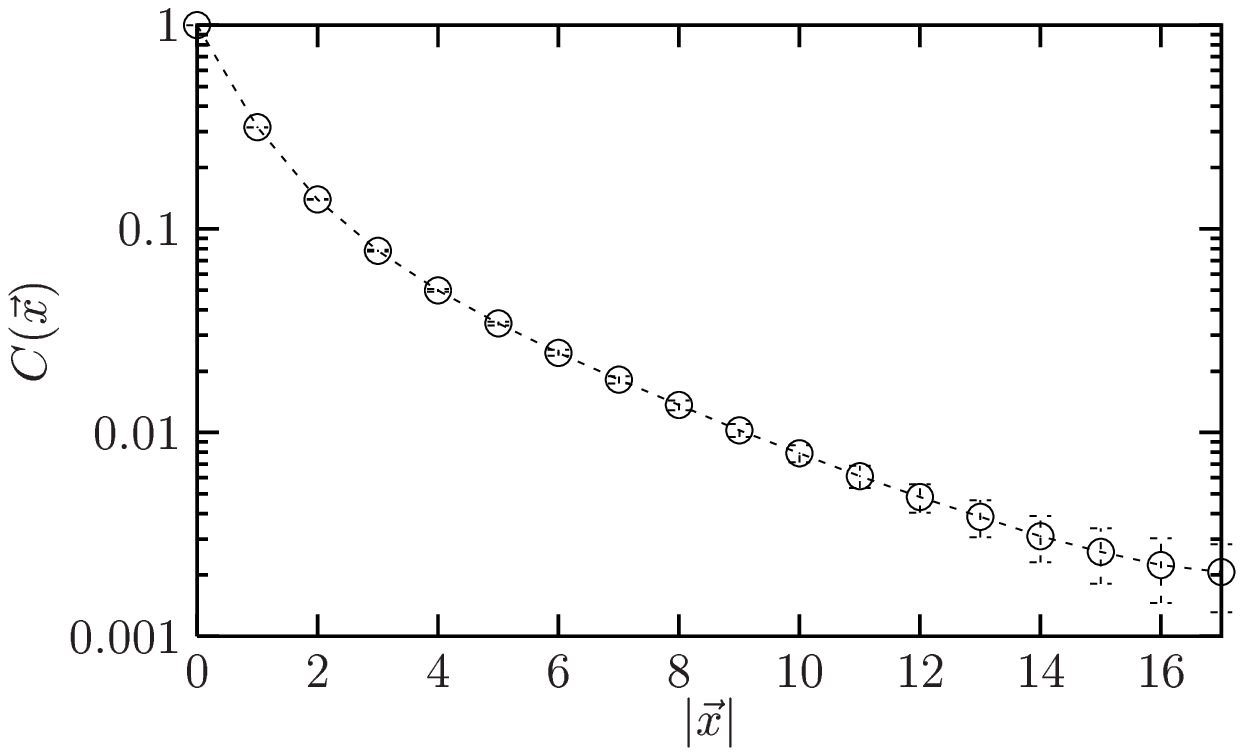,width=.45\linewidth}}%
  \hspace*{5mm}
\subfigure[\footnotesize{$\lambda = 2,~m^{2}=-0.3$}]
{\epsfig{figure=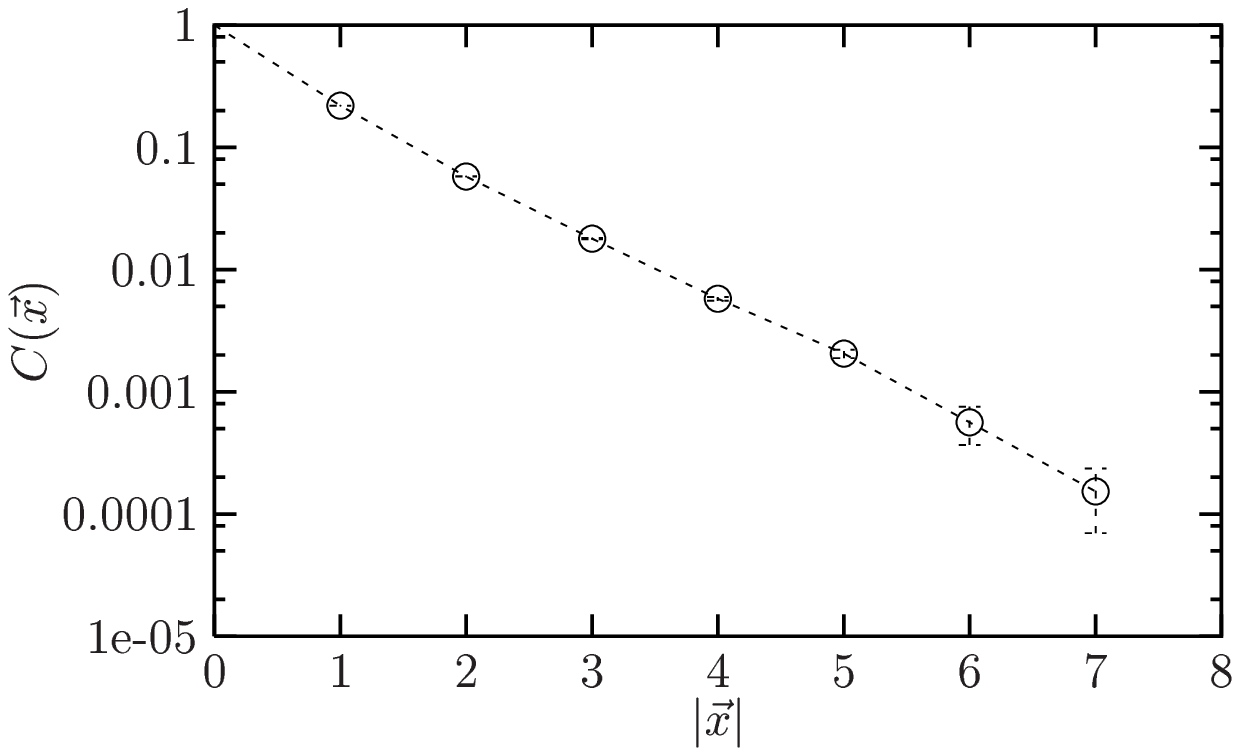,width=.45\linewidth}}
\caption{{\it Decay of the spatial correlator $C(\vec x)$ 
defined in eq.\ (\ref{CC}) at $N=35$ in the
disordered phase, close to the ordering phase transition.
On the left we see a clear deviation from the exponential decay.
At larger $\lambda$ (on the right) the standard exponential behavior 
is approximated.}}
\label{decay}
\vspace*{-2mm}
\end{figure}

Next we consider the spatial correlator at a fixed time,
\be  \label{CC}
C(\vec x ) = \la \phi (\vec 0 ,t) \phi (\vec x , t) \ra \ .
\ee
As it can be expected, in the disordered phase this decay is fast, 
but in contrast to our daily experience 
in the commutative world, it is in general not exponential
(nor polynomial),
as Fig.\ \ref{decay} (on the left) shows. It may be
plausible that the spatial decay behavior is somehow distorted by 
the NC geometry; what is perhaps more surprising
is that 
the decay moves closer
to an exponential as $\lambda$ increases (so that the NC effects are
expected to be enhanced), see Fig. \ref{decay} (on the right).
This observation might be explained in the light of the pole
structure described in Ref.\ \cite{MRS}.

\begin{figure}[htbp]
 \centering
  \subfigure[\footnotesize{$\lambda = 0.6,~m^{2}=-0.7$}]
{\epsfig{figure=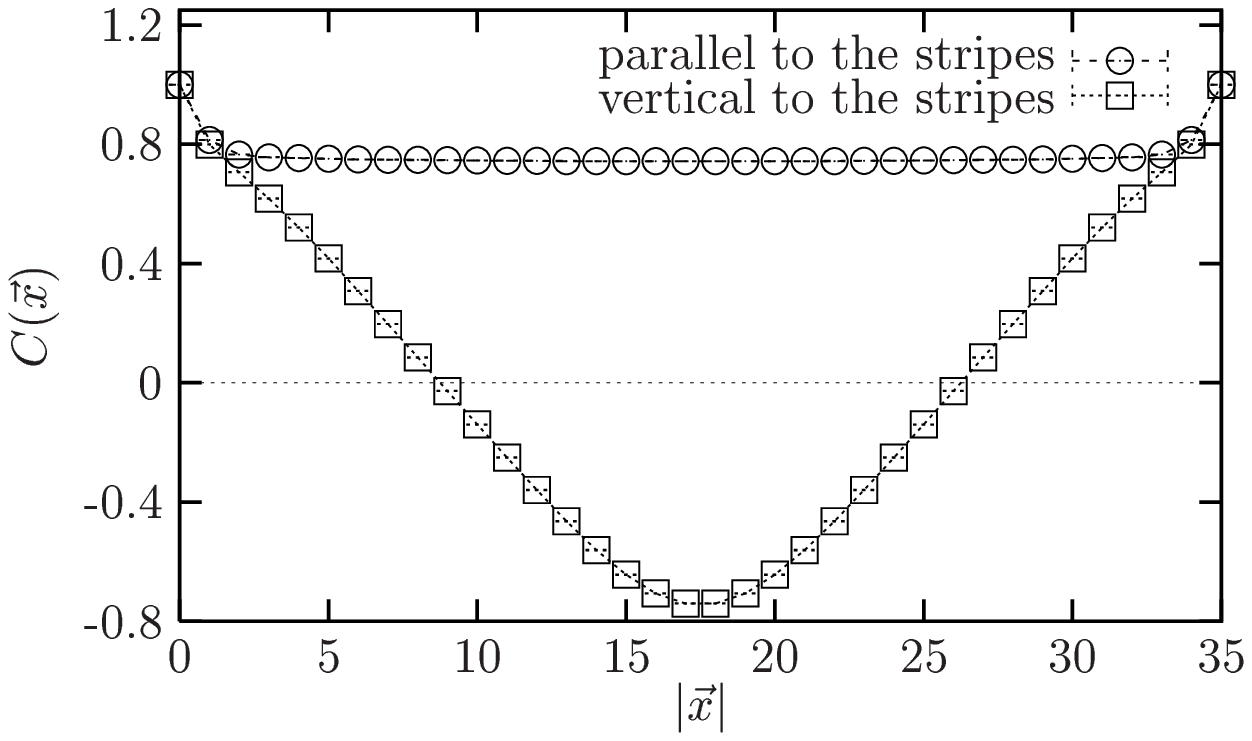,width=.45\linewidth}}%
  \hspace*{5mm}
\subfigure[\footnotesize{$\lambda = 6,~m^{2}=-4$}]
{\epsfig{figure=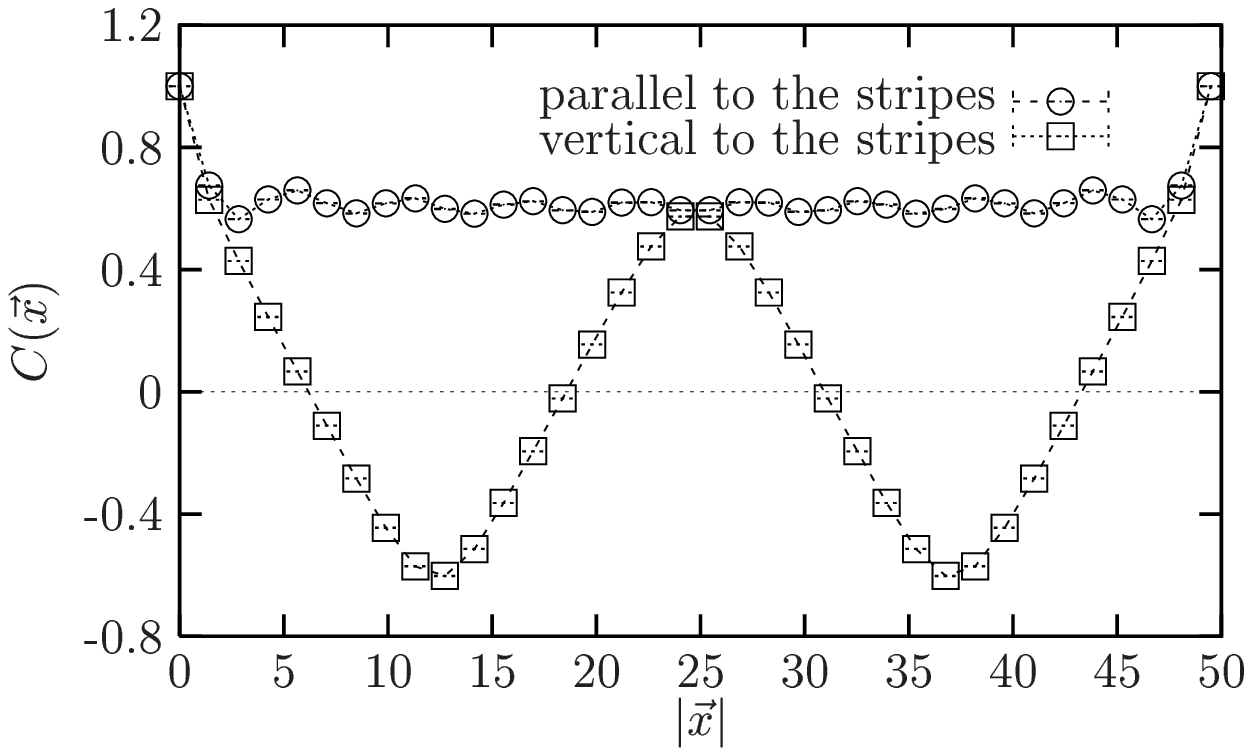,width=.45\linewidth}}
\caption{{\it The profile of $C(x_{1},0)$ and $C(0,x_{2})$ at $N=35$ 
for two stripes parallel to an axis (left) and two diagonal
stripes, as they occur at stronger coupling (right).}}
\label{Cstripes}
\vspace*{-2mm}
\end{figure}

Turning now to the striped phase, the pattern with stripes
parallel to one of the axes can be visualized well
by plotting $C(x_1 ,0)$ and $C(0, x_2 )$. 
In Fig.\ \ref{Cstripes} we show two examples where the cuts
through a configuration of two stripes parallel to an axis
(on the left) resp.\ two diagonal
stripes (on the right) can easily be recognized.

\begin{figure}[htbp]
 \centering
  \subfigure{\epsfig{figure=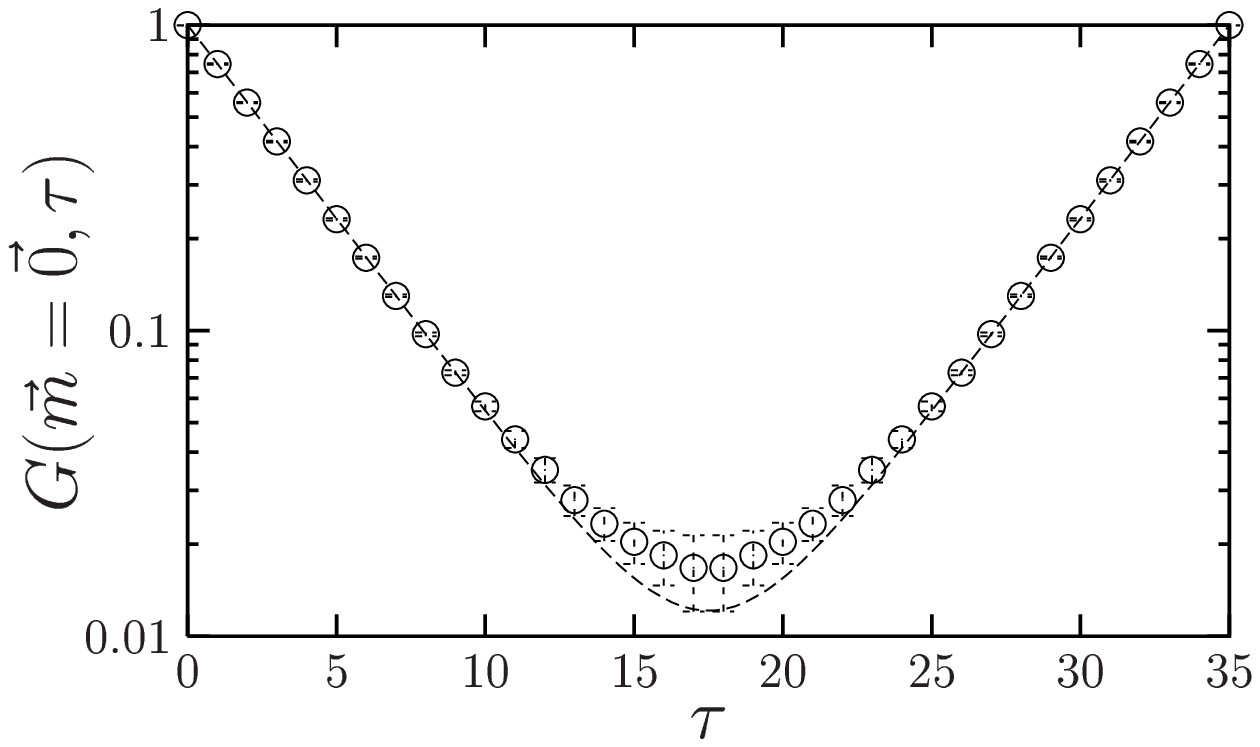,width=.45\linewidth}}%
  \hspace*{5mm}
\subfigure
{\epsfig{figure=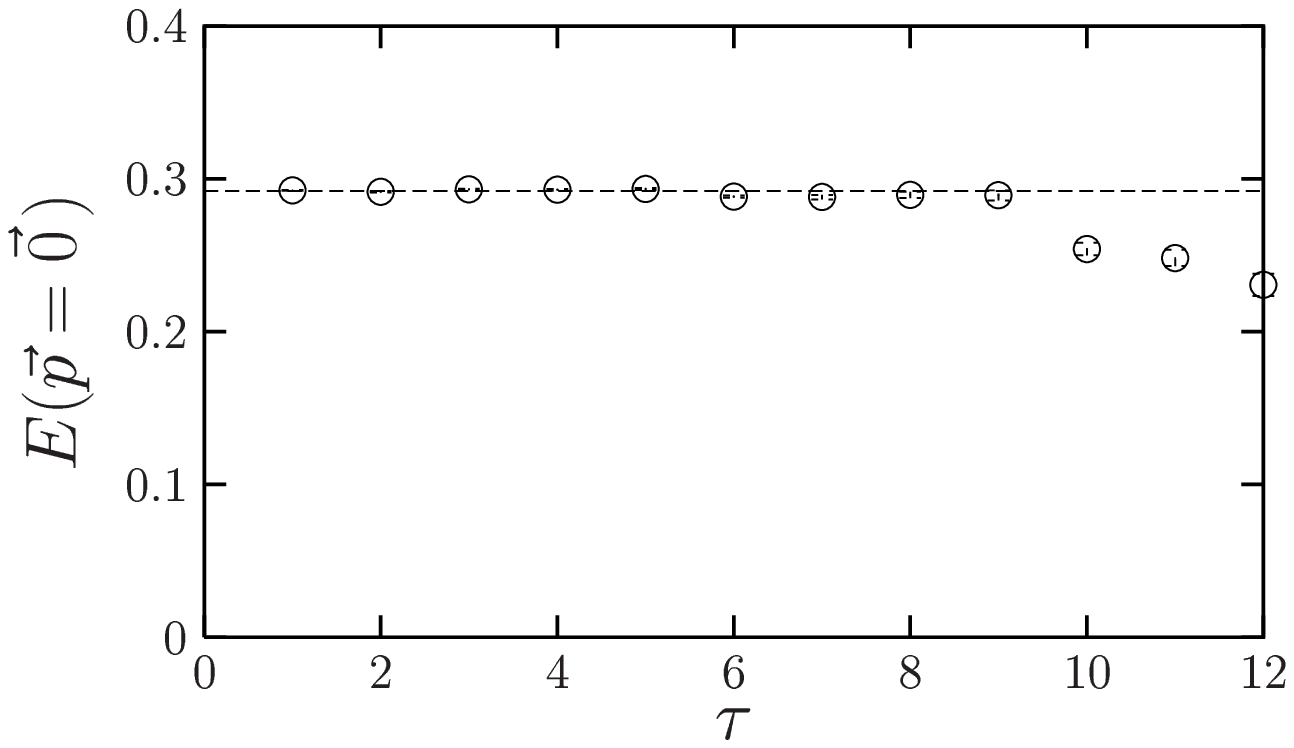,width=.45\linewidth}}
\caption{{\it On the left: the temporal correlator $G(\tau )$,
defined in eq.\ (\ref{Gtau}), at $N=35$, $\lambda = 0.29$, 
$m^{2} = -0.11$. On the right we show the rest energy extracted
at different values of $\tau$ according to eq.\ (\ref{energy}),
which has a clear plateau.}}
\label{Tdecay}
\vspace*{-2mm}
\end{figure}

Now we want to consider the correlations in the direction of the 
Euclidean time. More precisely, we consider the correlators of
two fields $\tilde \phi (\vec p ,t)$ at the same momentum $\vec p$ 
but at different times.
For instance, Fig.\ \ref{Tdecay} on the left shows
\be  \label{Gtau}
G (\tau ) := \la \tilde \phi (\vec p = \vec 0 , t ) 
\tilde \phi (\vec p = \vec 0 , t + \tau ) \ra
\ee
in the disordered phase, close to the phase transition.
Since the time direction is commutative, we rather expect here
the usual behavior, and indeed these correlators follow neatly
the exponential with periodic boundary conditions, i.e.\ a {\tt cosh}
shape. This shape allows us to extract the energy at rest
--- or effective mass --- in the exponentially dominated sector 
(not too close to the center) as
\be  \label{energy}
E (\vec p = \vec 0 ) = - \ln \frac{G(\tau + 1)}{G(\tau )} \ .
\ee
By varying $\tau$ we find a convincing plateau, see Fig.\ \ref{Tdecay}
on the right.

\begin{figure}[htbp]
 \vspace*{-3mm}
 \centering
  \subfigure[\footnotesize{$N=45,~ \lambda = 0.035,~m^{2}=-0.009$}]
{\epsfig{figure=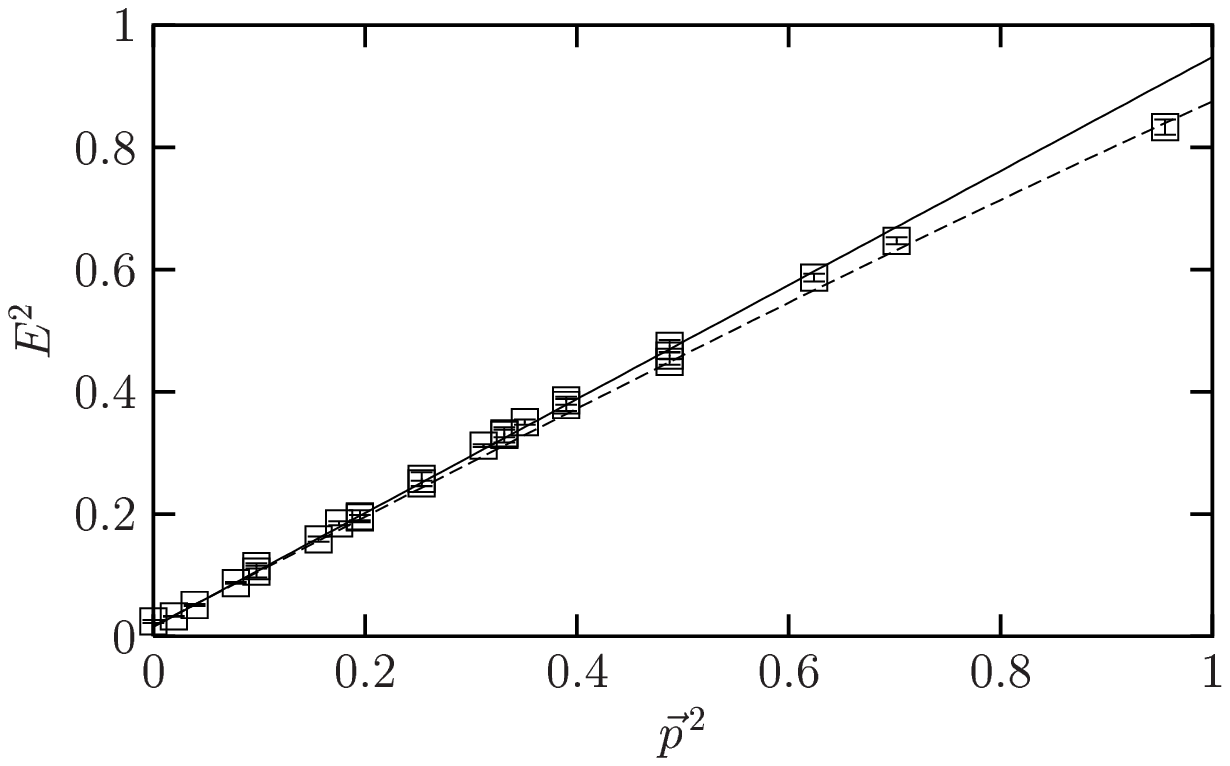,width=.45\linewidth}}%
  \hspace*{5mm}
\subfigure[\footnotesize{$N=55,~ \lambda = 55,~m^{2}=-15$}]
{\epsfig{figure=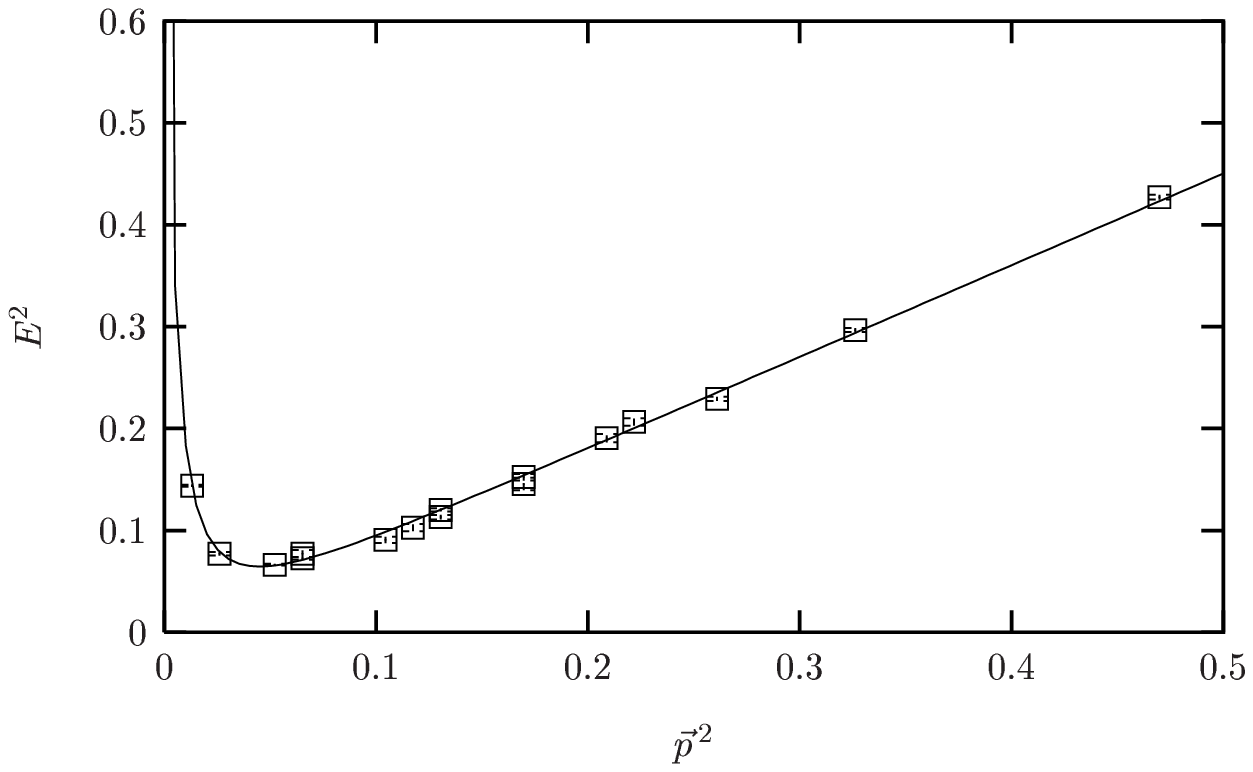,width=.45\linewidth}}
\caption{{\it The dispersion relation $E^{2}(\vec p^{\, 2})$ in the 
disorder phase, close to the ordering transition.
On the left: at weak coupling --- above the uniform phase ---
it takes the standard shape known from the commutative world,
including the lattice correction (dashed line).
On the right: at strong coupling --- above the striped phase ---
the rest energy increases and the energy minimum moves to finite
momenta. The line is a four parameter fit to the one loop prediction
\cite{diss,lat03}. Although we are still in the disordered phase, the 
system already feels the trend towards a stripe formation.}}
\label{disp}
\vspace*{-2mm}
\end{figure}

\begin{figure}[htbp]
 \begin{center} 
   \includegraphics[width=0.7\linewidth]{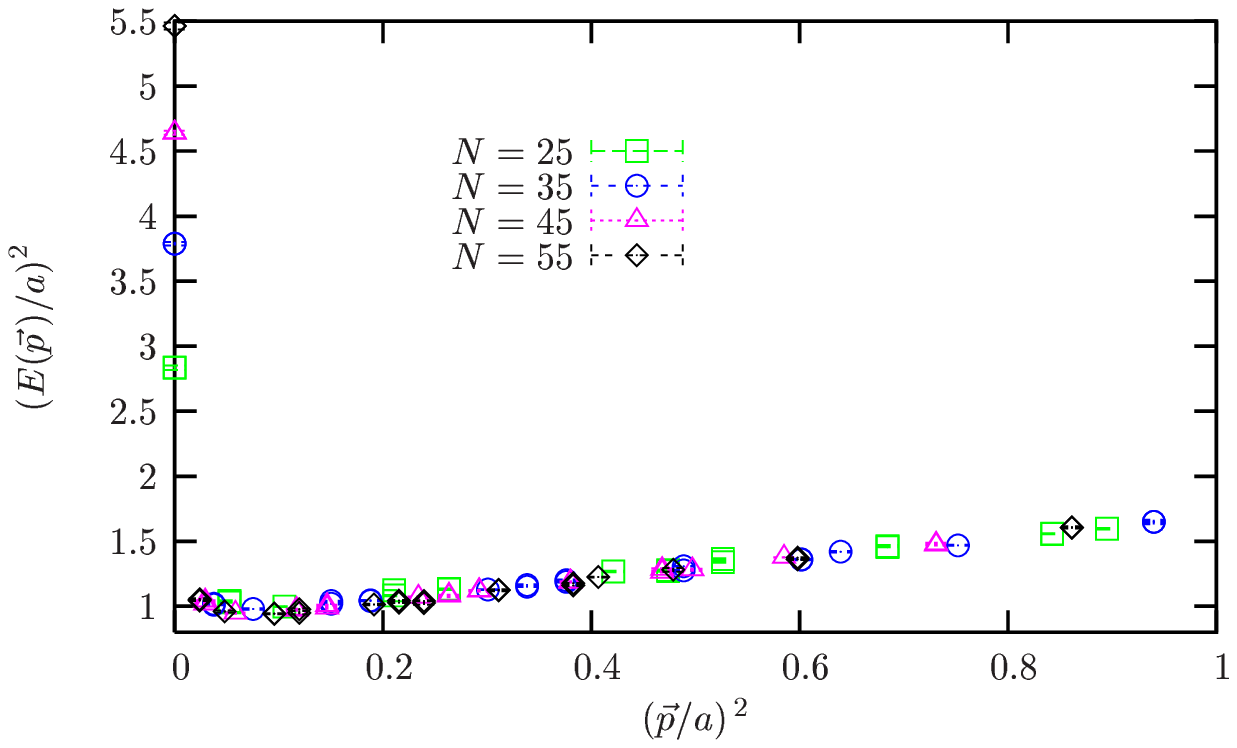} \\
 \end{center}
\caption{{\it The dispersion relations at different sizes $N$ and
physical lattice spacings $a$. The product $N a^{2} = 100$ is kept
constant, which corresponds to the prescription of the double scaling
limit. We see that the rest energy seems to diverge at
large $N$, whereas the shape of the
rest of the dispersion relation stabilizes.}}
\label{multidisp}
\end{figure}

We can now repeat this procedure also at non-vanishing momenta $\vec p$,
which finally yields the full dispersion relation $E( \vec p )$.
We still stay in the disordered phase, close to the ordering transition.
If we are close to the uniform order (small $\lambda$), the dispersion
has its usual linear shape (up to finite size corrections)
as in the commutative case, see Fig.\ \ref{disp} on the left.
However, the shape changes if we move to the vicinity of the
striped phase, see Fig.\ \ref{disp} on the right. Now
the rest energy increases, since the effects of UV/IR mixing set in.
In the large $N$ double scaling limit it is expected to diverge. 
The minimum 
moves to a finite value of the momentum, which corresponds to
the stripe patterns shown in Fig.\ \ref{snap}.
If we enlarge the lattice, we expect the dispersion relation
to stabilize if the axes are taken in physical units
(an exception is of course the close vicinity of $\vec p = \vec 0$).
The required re-scaling of the axes determines the physical
lattice spacing $a$. A clear hint that such a stabilization can 
indeed be achieved is the preliminary Fig.\ \ref{multidisp} --- 
for the corresponding definition of the $a$ (which also stabilizes
the spatial correlator) we refer to Ref.\ \cite{prep}.
We see here in particular that the energy minimum stabilizes
in physical units, which means that at large $N$ we find an infinite
number of stripes with a stable average width, which can be denoted
as a {\em zebra pattern}. (Of course this also includes the interference
of stripes in several directions; this is what we meant by checker-field
pattern.) Once this is fully demonstrated, we have 
established the continuum limit and the final confirmation of
the existence of the striped phase, as conjectured by Gubser and
Sondhi \cite{GubSon}.\\

{\small 
We finally remark that the occurrence of stripes in the lattice formulation
was also observed in the case of two NC dimensions in $d=2$
\cite{aristocats,BHN,diss}
and in $d=4$ \cite{prep}. In both cases, this observation is not
trivial from the theoretical point of view: since such a phase 
breaks the translation invariance spontaneously, one may wonder
if it can possibly exist in $d=2$ at all \cite{GubSon}, 
c.f.\ Section 2. However,
there is no contradiction with the Mermin-Wagner Theorem because the
proof of the latter assumes locality and an IR regular behavior,
which is both not provided here.
In $d=4$, on the other hand, the situation might be special
because we are dealing with the critical dimension in view
of the renormalization group \cite{ChenWu}, and an exciting
question would be if there are any news about triviality
once $\Theta$ enters the game.}

\section{Conclusions}

We have presented results from a numerical study of the $\lambda \phi^{4}$
model in 3d Euclidean space, where the two spatial directions are
NC, whereas the Euclidean time is commutative. The system was first
lattice discretized and then mapped onto a dimensionally reduced
matrix model. On each time lattice point one obtains a Hermitian
$N \times N$ matrix, where $N \times N$ is the spatial lattice volume.
We denote the physical (i.e.\ dimensionful) lattice space as $a$.
The double scaling limit $N \to \infty$, at $N a^{2} = ~constant$
describes the limit of an infinite volume in the continuum
at a finite non-commutativity parameter $\theta$. This means
that the UV limit and the thermodynamic limit are intertwined
in this case, as a consequence of UV/IR mixing.

We find a phase diagram which stabilizes for $N \gsim 25$.
The ordered regime splits into a uniform and a striped phase,
which is consistent with the conjecture by Gubser and Sondhi.
We discussed the order parameter, which is able to detect
different types of stripe patterns.
The spatial correlators in the disordered phase decay in some
irregular way (fast but in general not exponential),
whereas the correlators in time direction --- at fixed spatial
momentum --- do decay exponentially. Based on the latter
property we can determine the dispersion relation, which reveals
the UV/IR mixing again. At large $\lambda$ the energy minimum 
drifts away from zero, in accordance with the occurrence of stripe 
patterns in the ground state. The $\Theta$-deformed dispersion relation
at large $\lambda$ seems to be IR divergent. Hence the short-ranged 
non-locality has once again also long range manifestations.\\

As an {\bf outlook} we hope to identify a similar {\em dispersion relation 
for the photon}, which would be of immediate phenomenological interest, as
we outlined in Section 1.
We repeat that the experimental search for bounds on $\theta$ based 
on a possible frequency dependence of the speed of light
rely so far on a one loop calculation \cite{photon}
(with an uncertain behavior at higher loops). Therefore a non-perturbative
result would certainly be valuable. A first step in that direction
was the study of Wilson loops in the 2d NC $U(1)$ gauge theory 
\cite{U1}.\\

{\bf Acknowledgement} \ \ It is a pleasure to thank J. Ambj\o rn, 
S. Catterall, F. Iachello,
D. L\"{u}st, Y. Makeenko and R. Szabo for useful
comments. The computations were performed at the PC clusters
at Humboldt Universit\"{a}t and Freie Universit\"{a}t, Berlin.

\end{document}

%% file: diagram.tex
\begin{figure}[htbp]
  \centering
  \unitlength=1.00mm \linethickness{0.4pt}
  \begin{picture}(120.00,20.00)
    \thinlines
    \put(9.00,0.00){\line(1,0){30.00}}
    \put(17.00,0.00){\vector(1,0){0.50}}
    \put(24.00,7.20){\circle{20.00}}
    \put(24.00,14.2){\vector(1,0){.5}}
    \put(12.00,3.00){\makebox(0,0)[l] {$\scriptstyle{p}$}}
    \put(34.00,6.00){\makebox(0,0)[l] {$\scriptstyle{k}$}}
    \put(17,-5){planar}

    \put(60.00,7.20){\line(1,0){18.00}}
    \put(65.00,7.20){\vector(1,0){.50}}
    \put(85.00,7.20){\line(1,0){8.00}}
    \put(75.00,7.20){\circle{20.00}}
    \put(75.00,14.2){\vector(1,0){.5}}
    \put(63.00,10.00){\makebox(0,0)[l]{$\scriptstyle{p}$}}
    \put(82.00,14.00){\makebox(0,0)[l]{$\scriptstyle{k}$}}
    \put(65,-5){non--planar}

  \end{picture}\\
  \vspace{.5cm}
  \caption{{\it The planar and non-planar one loop contribution to the
two--point function (\ref{2point}), given in eqs.\ (\ref{2point1loop}).}}
  \label{UVIRfig}
\end{figure}

%% file: krakow.bbl
\begin{thebibliography}{50}

\bibitem{Sny} H.S. Snyder, Phys. Rev. 71 (1947) 38.

\bibitem{CNY} C.N. Yang, Phys. Rev. 72 (1947) 874.

\bibitem{Connes} A. Connes and M. Rieffel,
Contemp. Math. 62 (1987) 237.

\bibitem{string} A. Connes, M.R. Doublas and A. Schwarz,
JHEP 02 (1998) 003. \\
N. Seiberg and E. Witten, JHEP 09 (1999) 032.     

\bibitem{ACNY} A. Abouelsaood, C. Callan, C.R. Nappi and S.A. Yost,
Nucl. Phys. B280 (1987) 599.

\bibitem{Girvin} S.M. Girvin and A.H. MacDonald, Phys. Rev. Lett.
58 (1987) 1252. \\
L. Susskind, {\tt hep-th/0101029}.\\
A.P. Polychronakos, JHEP 06 (2001) 070.

\bibitem{Barbon} J.L.F. Barbon, ``Introduction to noncommutative theories'',
Lecture at ICTP Trieste, Summer 2001.

\bibitem{DFR} S. Doplicher, K. Fredenhagen and J.E. Roberts,
Phys. Lett. B331 (1994) 39; 
Commun. Math Phys. 172 (1995) 187.

\bibitem{causal} N. Seiberg, L. Susskind and N. Toumbas,
JHEP 0006 (2000) 044.\\
H. Bozkaya, P. Fischer, H. Grosse, M. Pitschmann, V. Putz,
M. Schweda and R. Wulkenhaar, Eur. Phys. J. C29 (2003) 133.

\bibitem{Kamel} G. Amelino-Camelia, L. Doplicher, S. Nam and Y.-S. Seo, 
Phys. Rev. D67 (2003) 085008.
                                             
\bibitem{photon} A. Matusis, L. Susskind and N. Toumbas,
JHEP 0012 (2000) 002.

\bibitem{Filk} T. Filk, Phys. Lett. B376 (1996) 53.

\bibitem{MRS} S. Minwalla, M. van Raamsdonk and N. Seiberg,
JHEP 02 (2000) 020.

\bibitem{Szabo} R.J. Szabo, Phys. Rept. 378 (2003) 207.

\bibitem{GubSon} S.S. Gubser and S.L. Sondhi, Nucl. Phys. B605 (2001)
395.

\bibitem{ChenWu} G.-H. Chen and Y.-S. Wu, Nucl. Phys. B622 (2002) 189.

\bibitem{AMNS} J. Ambj\o rn, Y.M. Makeenko, J. Nishimura and R.J.
Szabo, JHEP 05 (2000) 023.

\bibitem{AIIKKT} H. Aoki, N. Ishibashi, S. Iso, H. Kawai, Y. Kitazawa and
T. Tada, Nucl. Phys. B565 (2000) 176.

\bibitem{TEK} A. Gonz\'{a}lez-Arroyo and M. Okawa,
Phys. Lett. 120B (1983) 174; Phys. Rev. D27 (1983) 2397.
 
\bibitem{BHN} W. Bietenholz, F. Hofheinz and J. Nishimura,
Nucl. Phys. B (Proc. Suppl.) 119 (2003) 941;
Fortsch. Phys. 51 (2003) 745.

\bibitem{diss} F. Hofheinz, Ph.D. Thesis, Humboldt Univ. Berlin, 2003.

\bibitem{aristocats} 
J. Ambj\o rn and S. Catterall,  Phys. Lett. B549 (2002) 253.

\bibitem{solid} D. Mihailovic and V.V. Kabanov, Phys.
Rev. B63 (2001) 054505.

\bibitem{lat03} W. Bietenholz, F. Hofheinz and J. Nishimura,
{\tt hep-th/0309182}.

\bibitem{prep} W. Bietenholz, F. Hofheinz and J. Nishimura,
in preparation.

\bibitem{U1} W. Bietenholz, F. Hofheinz and J. Nishimura,
JHEP 09 (2002) 009.

\end{thebibliography}
